\documentclass{article}
\pdfoutput=1
\usepackage{graphicx}

\def\beq{\begin{eqnarray}}
\def\eeq{\end{eqnarray}}
\def\bea{\begin{eqnarray*}}
\def\eea{\end{eqnarray*}}

\def\centeron#1#2{{\setbox0=\hbox{#1}\setbox1=\hbox{#2}\ifdim
\wd1>\wd0\kern.5\wd1\kern-.5\wd0\fi
\copy0\kern-.5\wd0\kern-.5\wd1\copy1\ifdim\wd0>\wd1
\kern.5\wd0\kern-.5\wd1\fi}}
\def\ltap{\;\centeron{\raise.35ex\hbox{$<$}}{\lower.65ex\hbox{$\sim$}}\;}
\def\gtap{\;\centeron{\raise.35ex\hbox{$>$}}{\lower.65ex\hbox{$\sim$}}\;}

\def\singleandthirdspaced{\baselineskip=\normalbaselineskip\multiply
    \baselineskip by 130\divide\baselineskip by 100}

\newcommand{\newc}{\newcommand}
\newc{\qbar}{{\overline q}}
\newc{\Kahler}{K\"ahler }
\newc{\deltaGS}{\delta_{\rm GS}}

\newcommand{\vev}[1]{\langle #1\rangle}

\begin{document}
\begin{titlepage}
\begin{flushright}
{\large arXiv:1212.4371 \\
SCIPP 12/15\\
}
\end{flushright}

\vskip 1.2cm

\begin{center}

{\LARGE\bf Discrete R Symmetries and Anomalies}

\vskip 1.4cm

{\large  Michael Dine and Angelo Monteux}
\\
\vskip 0.4cm
{\it Santa Cruz Institute for Particle Physics and
\\ Department of Physics,
     Santa Cruz CA 95064  } \\
\vskip 4pt

\vskip 1.5cm

\begin{abstract}
We comment on aspects of discrete anomaly conditions  focussing particularly on
$R$ symmetries.   
We review the Green-Schwarz cancellation of discrete anomalies,
providing a heuristic explanation why, in the heterotic string, only the
``model-independent dilaton" transforms non-linearly under discrete symmetries; this argument suggests that, in other
theories, multiple fields might play a role in anomaly cancellations, further weakening any anomaly
constraints at low energies.  We provide examples in open string theories
of non-universal discrete anomalies at low energies. We then consider the fact that $R$ symmetries
are necessarily broken at low energies.  We exhibit dynamical models, 
in which fields charged under the Standard Model gauge group (for example, a doublet and a triplet) gain roughly equal masses, but where the doublet and the triplet possess different discrete charges and the low-energy anomaly conditions fail.
\end{abstract}
\end{center}

\vskip 1.0 cm

\end{titlepage}
\setcounter{footnote}{0} \setcounter{page}{2}
\setcounter{section}{0} \setcounter{subsection}{0}
\setcounter{subsubsection}{0}

\singleandthirdspaced

\section{Introduction}

It is conceivable that discrete symmetries play an important role in low energy physics.  They are an important tool for model builders.
Starting with the work of Krauss and Wilczek\cite{krausswilczek}, it has been argued that such symmetries should be gauge symmetries,
and as such should be free of anomalies.  A set of consistency conditions were formulated by Ibanez and Ross\cite{ibanezross}.  Subsequently,
it was noted that in heterotic string theories, only a weaker set of conditions hold\cite{banksdinediscrete}, and that 
\begin{enumerate}
\item  Anomaly conditions can only be applied with respect to non-abelian symmetries
\item  Anomalies can be compensated by a Green-Schwarz mechanism.  This is only meaningful in weak coupling (more precisely in
certain extremes of the moduli space, where the breaking terms are exponentially suppressed).
\end{enumerate}
Various authors have enforced various rules on model building ranging from the stronger Ibanez-Ross constraints
to the weaker ones of \cite{banksdinediscrete}\footnote{An early example, which finds striking constraints, appears in \cite{yanagidaanomalies}.}.  But even the latter are arguably too strong; this will be the subject of the
present paper.  First, these constraints are frequently applied to discrete $R$ symmetries.  But, given the small size of the cosmological
constant, $R$ symmetries are necessarily broken at some high energy scale.  Fields may gain mass as a result of this breaking,
negating  any constraint.  We will illustrate this effect with explicit models for dynamical $R$ symmetry
breaking.  Second, we will see that studies of anomalies in the heterotic string are limiting.  For the compactifications which have
been considered to date, there is a simple argument why the anomalies are universal.  We will (re)consider a set of Type II orientifold
models\cite{ibanezuranga,dinegraesser}, and see that these have non-universal low energy anomalies.\footnote{In
\cite{ibanezuranga} discrete symmetries are the remnants of the breaking of a gauge symmetry and because of that they are anomaly-free at low and high energy. This is not the most general way of getting a discrete symmetry: in the Type II examples discussed here the discrete symmetry has a geometric origin, that is, it is due to the compactification of the extra dimensions.
}
After discussing gauge unification, we conclude from these observations that there are no compelling reasons, absent a specific and detailed microscopic theory, to impose discrete anomaly constraints at low energies.

In the next section, we review briefly the situation with anomalies in the heterotic theory.  In section \ref{multiple}, we explain why
in many compactifications of the heterotic string the anomalies are universal.  In section \ref{typeii}, we 
examine several Type II models which provide examples of non-universal anomalies.  Then we discuss breaking of $R$ symmetries in 
sections \ref{rbreaking} and \ref{rbreakingmodels}, noting
that fields can gain mass at a variety of scales, and consider subtleties of possible constraints from coupling unification. 
We remark about implications for model building in our concluding section.


\section{Anomaly Conditions in Heterotic String Theory}
\label{heterotic}

In a unified field theory, it is simple to argue that if the high energy theory is anomaly free, so is the low energy theory.  Whatever
fields gain mass as a consequence of the unified symmetry breaking, they are necessarily
vector-like with respect to the unbroken symmetries.  On the other hand, the usual 't Hooft
type of argument would not seem to apply in this case; discrete symmetries could exhibit anomalies with respect to
$U(1)$'s without leading to any obvious inconsistency.  Instead, we can view the cancellation of anomalies as a simple
algebraic fact.
If anomalies are cancelled by a Green-Schwarz mechanism, in a simple group, the anomalies of the low energy theory
must be equal, by the same algebraic argument as in the case without a GS cancellation.  

In string theory, one does in fact find that the only anomaly conditions which hold, in general, are those
involving non-abelian gauge symmetries.  Moreover,
it has been known for some time that in string theories, anomalies in discrete symmetries may be cancelled by a Green-Schwarz mechanism.
In other words, there may be a (pseudo) modulus, $\Phi$, transforming under the discrete symmetry as $\Phi \rightarrow \Phi + {2 \pi i \over N}$
(for the case of a $Z_N$ symmetry) and with coupling $\Phi W_\alpha^2$. 
From a low energy perspective, once one has allowed for the possibility of Green-Schwarz cancellations, it would seem that there
are no constraints on discrete anomalies, provided that one has sufficient numbers of light pseudo moduli with suitable
couplings to $W_\alpha^2$.\footnote{Indeed, one might wonder whether the existence of pseudo moduli is a requirement, with
anomalies instead being cancelled by heavy fields with multiple ground states related by the discrete symmetry transformation.
While logically possible, we have found it difficult to model this phenomenon.}  These moduli, if sufficiently weakly coupled, might play no other significant 
role in low energy physics (cosmology aside).  So there is no a priori reason, from purely macroscopic considerations, to impose anomaly constraints.
On the other hand, in the heterotic string, at least in all examples which have been studied to date, only one modulus transforms under the discrete symmetries, 
and so all would-be anomalies are identical.


Both of these issues are nicely illustrated by the $O(32)$ heterotic string.  We can consider the textbook\cite{gsw} example
of compactification of the theory on a Calabi-Yau manifold described by the vanishing of a quintic polynomial in $CP^4$.
In terms of the coordinates of $CP^4$, $x_i$, $i=1,\dots 5$,
\beq
P = \sum x_i^5 = 0.
\eeq
This theory exhibits a $Z_5^4 \times S_5$ symmetry.  We mod out by the transformation
\beq
x_i \rightarrow \alpha^{i} x_i;~~\alpha= e^{2 \pi i \over 5},
\eeq
which is a freely acting symmetry.  This reduces the number ``generations" ($26$'s of $O(26)$ with charge $+1$) from $101$ to
$21$; there is still one anti generation.  We combine this with the action of a Wilson line.  We take the Wilson line to lie in a $U(13)$
subgroup of $O(26)$:
\beq
U = {\rm diag} \left (\alpha^{k_1},\dots ,\alpha^{k_{13}} \right ).
\eeq
The requirement of modular invariance is
\beq
{1 \over 2N^2} \sum (k_i^2 - k_i N) = {n \over 5}\,, \qquad N=5
\eeq
for some integer $n$.

Consider a particular choice of $U$:
\beq
k_i = 1, ~i=1 \dots ,9;~~~k_i =2,~ i = 10,\dots,13.
\eeq
In this case, the low energy group includes $SU(9) \times SU(4) \times U(1)  \times U(1)$.  To investigate
the question of anomalies, we choose one of the surviving symmetries:
\beq
x_1 \rightarrow \alpha x_1.
\eeq
It is a simple exercise to determine the anomalies.
For $SU(4)$, one finds
$\alpha^8 = \alpha^3$, while for $SU(9)$ one finds $\alpha^{18} = \alpha^3$.  (The counting, here, is particularly simple.
The matter fields include $20$ fields in the $(9,1)$ and $20$ in the $(1,4)$.  The $R$ charges are easily worked out.  The
product of scalar $R$ charges, in each case, is $1$.  The fermion and scalar charges differ by a fifth root of unity,
which cancels out for the 20 fields in the instanton determinant.  This just leaves the $8$ and $18$ gaugino
zero models for the two groups.)  However, the discrete symmetry would appear
to have no anomaly with respect to the $U(1)$'s, since the anomalies, again, cancel among the matter fields, but now the
gauginos are neutral.

Other choices of Wilson line similarly illustrate these phenomena of Green-Schwarz cancellations, as well as
the lack of a constraint for the anomalies in the symmetry relative to $U(1)$ gauge groups.

\section{On the Possibility of Multiple Green-Schwarz Cancellations}
\label{multiple}

In the heterotic string, discrete anomalies have been studied in a range of Calabi-Yau, orbifold (symmetric and asymmetric),
and other compactifications\cite{banksdinediscrete,dinegraesser}.  In these compactifications, one finds that
all anomalies can be cancelled by assigning a transformation law to the dilaton of the weak
coupling theory, i.e. the anomalies are all equal (again this refers to anomalies with respect to non-abelian
gauge groups).  Many workers have concluded from these observations that this is a general requirement
which should be imposed on low energy theories.

But a priori this is not obvious.  In string models, in particular, there are typically multiple moduli fields.
Generically, several if not all of these will couple to some of the various gauge groups (i.e. they will
exhibit couplings of the form $X W_\alpha^2$), and, if these couplings differ, and if the fields have
different transformation laws under the discrete symmetries, the anomalies could
differ.  We now argue that the heterotic string theory is special; that this is not a result which need
hold in a general theory coupled to gravity.

Consider, for example, the case of the quintic.  For special values of the radius, the theory is known
to exhibit an enhanced symmetry\cite{gepner}.  Consider the theory at such a point.  If we wish to assign
to the radial dilaton a transformation property under the $Z_5$, this transformation is necessarily
linear (a non-linear transformation law, for example, would not be consistent with the enhanced gauge symmetry),
but perturbatively the dilaton is invariant.
Any fields which gain mass as one moves away from this point are necessarily in vector-like representations of the
unbroken symmetries.

This behavior would
appear, in the case of the heterotic string, to be generic.  For example, toroidal and orbifold compactifications
typically have points in their moduli spaces where all of the moduli, apart from the dilaton, are charged.
The same is true of Calabi-Yau compactifications that admit a Gepner description
somewhere on their moduli space.  On the other hand, the space of $N=1$ compactifications
of string theory is larger than these particular compactifications of the heterotic string, in which case it is possible that there are moduli
which do not experience such enhanced symmetries, which could also play a role in anomaly cancellation.
So it is worthwhile to study a broader class of theories.

\section{Non-Universal Anomalies in Type II Theories}
\label{typeii}

A clue as to where to look for non-universal anomalies is provided by the study of {\it continuous} $U(1)$ symmetries
in string models. In \cite{ibanezuranga} 
Ibanez and Uranga studied Type II orientifolds and open strings, and found that
often there are several anomalous $U(1)$'s, and that these anomalies
are {\it not} universal, the anomalies being cancelled by
various axion-like fields.  Related to these anomalies, there are also Fayet-Iliopoulos terms, controlled by
these various moduli.  These may vanish at certain points on the moduli
space.

In addition to $U(1)$'s, many of these models have discrete symmetries as well, with assorted anomalies, as discussed in \cite{dinegraesser}.
In that reference, it was noted that the ratios of the discrete anomalies with respect to the non-Abelian groups were the same as the ratios of the continuous anomalies with respect to the same groups (modulo suitable integers). Then, one could define a non-anomalous discrete symmetry by combining the discrete symmetry with the $U(1)$: denoting by $\alpha$ the parameter of the continuous $U(1)$, the discrete symmetry would be $Z_N'=Z_N\times U(1){_{\alpha=\frac{2\pi}N}}$.  But this is not really the interesting question.  Instead, one can ask whether, once one has accounted for the
Fayet-Iliopoulos terms, there are surviving discrete symmetries at low energies, whether these are sometimes anomalous, and
whether, if so, these anomalies are universal.  Alternatively, if the FI term  vanishes, the gauge field and the corresponding modulus are massive,
and one can study the remaining discrete symmetries in the low energy theory and their anomalies in isolation. In what follows, we will
give examples of both phenomena.

A simple example is provided by a $Z_3$ orbifold, developed in \cite{ibanezuranga} and studied in
\cite{dinegraesser}.  At a microscopic level, this theory has gauge group
\beq
SU(12) \times SO(8) \times U(1).
\eeq
It has two discrete $Z_6$ discrete $R$ symmetries (on a subspace of the moduli space).  The massless field content is:
\beq
2(12,8,1;\gamma^{1/2}) + (12,8,1;\gamma^{-1/2}) + 2 (\overline{66},1,-2; \gamma^{1/2}) +
 (\overline{66},1,-2; \gamma^{-1/2}) + (143,1,0;\gamma^{-1/2}) + (1,28,0;\gamma^{-1/2}).
 \eeq
 Here $\gamma = e^{2 \pi i \over 6}$, and the $\gamma$ quantum number
 refers to the transformation property of the fermionic component of the
 multiplet under the discrete
symmetry.  The last two sets of fields are the gauginos of $U(12)$ and $SO(8)$, respectively.
 There are a variety of anomalies.
 There are a set of moduli,  chiral fields, $\Phi_i$, $i=1,2$ associated with the twisted sectors.  These fields cancel the anomalies through their couplings to the gauge $U(1)$'s $W_\alpha^2$; they also give rise to Fayet-Iliopoulos D-terms.  
 These are given by:
 \beq
 D= 6 \sqrt{3}( \Phi_1 - \Phi_2).
 \eeq

If the $D$ term vanishes, the gauge bosons are massive, and there is no $U(1)$ in the low energy theory.
The discrete anomalies with respect to $SU(12)$ and $SO(8)$ are $\gamma^3$ and $1$, respectively, i.e.
they are not universal.
There is still one light
 linear combination of $\Phi_1$ and $\Phi_2$, $\Phi_1 + \Phi_2$.  This, along with the dilaton, cancels the discrete anomaly. 
 Using the analysis of \cite{ibanezuranga}, one can compute the couplings of the two twisted moduli to the gauge groups
 $SU(12)$ and $SO(8)$.  
 \beq
 f_{SU(12)} = S -6  (\Phi_1 + \Phi_2)
 ~~~~f_{SO(8)} = S + 4 (\Phi_1 + \Phi_2).
 \eeq
 $\Phi \equiv \Phi_1 + \Phi_2$ is light.  Both $S$ and $\Phi$ must transform under the discrete symmetry to cancel the anomaly.

 Consider the case that the $D$ term is non-zero and positive.  Then we can cancel the $D$ term by giving an expectation value to one
 of the 
 $(\overline{66},1,-2; \gamma^{1/2})$ fields,
 \beq
\langle \overline{66}\rangle = v \left (\begin{matrix}{ \sigma_2 &0 &0 & 0 & 0 & 0 \cr 0 & \sigma_2 & 0 & 0& 0& 0 \cr 0 & 0 & \sigma_2  & 0& 0& 0
 \cr  0 & 0 & 0 &\sigma_2 & 0 & 0 \cr 0 &0 &0 & 0 &\sigma_2 & 0 \cr 0 & 0 & 0 & 0 & 0 & \sigma_2 }\end{matrix} \right ).
 \eeq
 This breaks the gauge symmetry to $SP(6) \times SO(8)$.   The scalar component of the $\overline{66}$ is neutral under the original
 discrete $R$ symmetry, so this symmetry is unbroken.  At low energies, the discrete anomalies are non-universal.  An $SP(6)$ instanton
 violates the symmetry by $\gamma^3$, while an $SO(8)$ instanton does not violate the symmetry at all.  There are three light moduli remaining
 in the low energy theory, more than enough to cancel the discrete anomaly.
 
Alternatively, for the same positive choice of the $D$-term sign, the field
$(\overline{66},1,-2; \gamma^{-1/2})
$ can obtain an expectation value.  The low energy gauge group is the same, but the unbroken discrete symmetry is different.
 The scalar component of the $\overline{66}$ transforms with phase $\gamma^{-1}$ under the original $Z_6$ symmetry, so at low energies,
 the unbroken discrete symmetry is a combination of the microscopic $Z_6$ and a $U(1)$ transformation. However, the extra $U(1)$ does
 not contribute to the anomalies at low energies, and one again has  non-universal behavior, with phases $(\gamma^3,1)$.

If the $D$ term has the opposite sign, it can be cancelled by giving a vev to any one of the $(12,8,1;\gamma^{\pm1/2})$ fields. For example, we can have
\beq
\langle (12,8) \rangle = \left (\begin{matrix}{ v & 0 & 0 & 0 & 0 & 0 & 0 & 0 & 0 & 0 & 0 & 0 \cr0 & v & 0 & 0 & 0 & 0 & 0 & 0 & 0 & 0 & 0 & 0 \cr0 & 0 & v & 0 & 0 & 0 & 0 & 0 & 0 & 0 & 0 & 0 \cr0 & 0 & 0 & v & 0 & 0 & 0 & 0 & 0 & 0 & 0 & 0 \cr0 & 0 & 0 & 0 & v & 0 & 0 & 0 & 0 & 0 & 0 & 0 \cr0 & 0 & 0 & 0 & 0 & v & 0 & 0 & 0 & 0 & 0 & 0 \cr0 & 0 & 0 & 0 & 0 & 0 & v & 0 & 0 & 0 & 0 & 0 \cr0 & 0 & 0 & 0 & 0 & 0 & 0 & v & 0 & 0 & 0 & 0}\end{matrix} \right ).
\eeq
The low energy gauge group is $SO(8)$ and the discrete symmetry is anomaly free.

 Similar results hold for various vacua of the models studied in \cite{dinegraesser}.  So non-universality seems to be the rule in the Type II constructions.
 One concludes from this that, from the perspective of a low energy model builder, there is no strong rationale for enforcing {\it any} discrete anomaly
 constraints.  Only if one is committed to some particular microscopic framework (e.g. heterotic strings compactified on orbifolds),
 or the assumption that there is no small parameter in the microscopic theory, can one justify
 such constraints.  

\section{Models with $R$ Breaking at Multiple Scales}
\label{rbreaking}

Even if one believes one has an underlying framework in which anomalies vanish or are universal, it is still not clear that one should
enforce anomaly constraints on $R$ symmetries.  This is because $R$ symmetries are necessarily broken at a high energy scale,
given the smallness of the observed cosmological constant.
As a result of this breaking, fields in non-vectorlike representations of the symmetry group may gain mass, even as the low
energy theories may (and often do) exhibit an  {\it approximate}  discrete $R$ symmetry.  The models of 
ref. \cite{dinekehayias}, for example, possess scalar fields whose vev's break discrete symmetries.  If they couple
to fields charged under the standard model, these fields gain mass; the low energy theory typically
still possesses an approximate $R$ symmetry, with
apparent anomalies, even if the microscopic theory was anomaly free.

One possible argument to impose anomaly constraints on $R$ symmetries arises from unification.  In the simple
models of \cite{dinekehayias}, all of the scalars transform in the same way under the $R$ symmetry, so in a unified model, one
might expect that complete multiplets gain mass, and that there would be no contribution to any anomaly from
these fields.  In this section, we present a model with scalars with different transformation properties,
in which fields in the color triplet and weak doublet representations gain comparable masses but also possess different transformation properties.

Suppressing dimensionless couplings, we take:
\beq\label{Rmodel}
W = S_1 \bar Q_f Q_f + S_1^3 + {S_2^2 \over M_p} \bar Q_a Q_a + {1 \over M_p} S_2^4.
\eeq
Here $f = 1,\dots, F_1$, $a = 1,\dots,F_2$ and the fields $Q_f,\ Q_a,$ are fundamental of an $SU(N)$.  
We want to integrate out the gauge fields and $Q,\bar Q$ fields and obtain an effective
superpotential for the singlets $S_1,~S_2$.  To do this we note, first, that for a set of quarks of mass $m_{f,f^\prime}$,
the effective superpotential at low energies (just the expectation value of the superpotential) is\cite{affleckdineseiberg}
\beq
W = (\det (m))^{1/N}\Lambda^{3-{N_f \over N}}
\eeq
where $\Lambda$ is the (holomorphic) renormalization group invariant scale of the underlying theory.
This follows from the flavor symmetries in the absence of the mass term, including the $R$ symmetry, and viewing $m$
as a spurion.  In our case, 
\beq
\det (m) = S_1^{F_1} S_2^{2 F_2} M_p^{-F_2}.
\eeq
Now we can obtain the vev's of $S_1$ and $S_2$ by finding the stationary points of the superpotential.
This can be done by straightforward algebra.  An alternative is to use symmetry principles. The expectation values of
$S_1$ and $S_2$ will be proportional to a power of $\Lambda$.  On the other hand, the theory has a non-anomalous
continuous $R$ symmetry under which $\Lambda$ transforms.  The $\Lambda$ transformation is determined by noting that
\beq
\Lambda = M_p e^{- \tau/b_0}
\eeq
where ${\rm Re} ~\tau = {8 \pi^2 \over g^2}; ~{\rm Im} \tau = i \theta$; $b_0=3N-F_1-F_2$.  Then, cancellation of anomalies under an $R$
transformation with parameter $\alpha$ requires
\beq
\tau \rightarrow \tau -2 i \alpha (N  - {1 \over 3} F_1 - {1 \over 2} F_2)
\eeq
Correspondingly, $\Lambda$ transforms as:
\beq
\Lambda \rightarrow \Lambda e^{2i \alpha \left ({N - {1 \over 3} F_1 - {1 \over 2} F_2 \over 3N - F_1 - F_2} \right )}.
\eeq
(as a check, one can repeat this argument for the pure gauge theory, to check that $\langle \lambda \lambda \rangle$
transforms with charge $2$).  So because $S_1$ transforms with $\beta^{2/3}$ and $S_2$ with phase
$\beta^{1/2}$ under an $R$ transformation $\beta=e^{i\alpha}$, their form is
\beq
\langle S_1 \rangle= M_p^{F_2 \over 2 F_1 + 3 F_2 - 6N} \Lambda^{2 F_1 + 2 F_2 - 6N \over 2 F_1 + 3 F_2 - 6N}  ~~~~
\langle S_2\rangle= M_p^{F_1 + 3 F_2 -3N \over 4 F_1 + 6 F_2 - 12N} \Lambda^{3 F_1 + 3 F_2 - 9N \over 4 F_1 + 6 F_2 - 12N} 
\eeq
These formulas agree with the straightforward algebra. 

\section{Models for Low Energy $R$ Anomalies}
\label{rbreakingmodels}

In the model of the previous section we took $S_1$ to transform with $\beta^{2/3}$, and $S_2$ with phase
$\beta^{1/2}$; anomaly freedom for the discrete symmetry yields:
\beq
\beta = e^{2 \pi i  {1 \over 6N - 2F_1 - 3 F_2}}
\eeq

Now we can couple $S_1$ and $S_2$ to fields charged under the Standard Model
gauge group.  As an example,
we introduce a field $q$, which is a color triplet, and a field $\ell$, a weak doublet. With a coupling of the form
\beq\label{SMcoupling}
 S_1\bar q q +S_2\bar\ell\ell,
\eeq
the fields $q$ and $\ell$ have $R$ charges $\frac23$ and $\frac34$, respectively. With given values of the parameters $N,\ F_1, \ F_2$ and $\Lambda$, the expectation values of $S_1$ and $S_2$ are set, and the fields $q$ and $\ell$ become massive.  If their masses are comparable, because they
have the structure of a complete $SU(5)$ multiplet, they will not spoil (and can even improve) gauge coupling unification\footnote{Unification of coupling,
with different $R$ charges for such fields, might arise, for example, in the sort of direct product representations discussed in \cite{wittendeconstruction}, or
in string theory models\cite{gsw}.}. 

If at high energies the $R$ symmetry was anomaly-free with respect to the SM non-abelian gauge groups, the low-energy theory will be anomalous. In the same way, if the high energy anomalies with respect to $SU(3)$ and $SU(2)$ were universal, they will no longer be in the low energy theory, as fields with different $R$ charges coupling to the different gauge groups have been integrated out.

\begin{figure}[tbhp]
\begin{center}
\includegraphics[width=13cm]{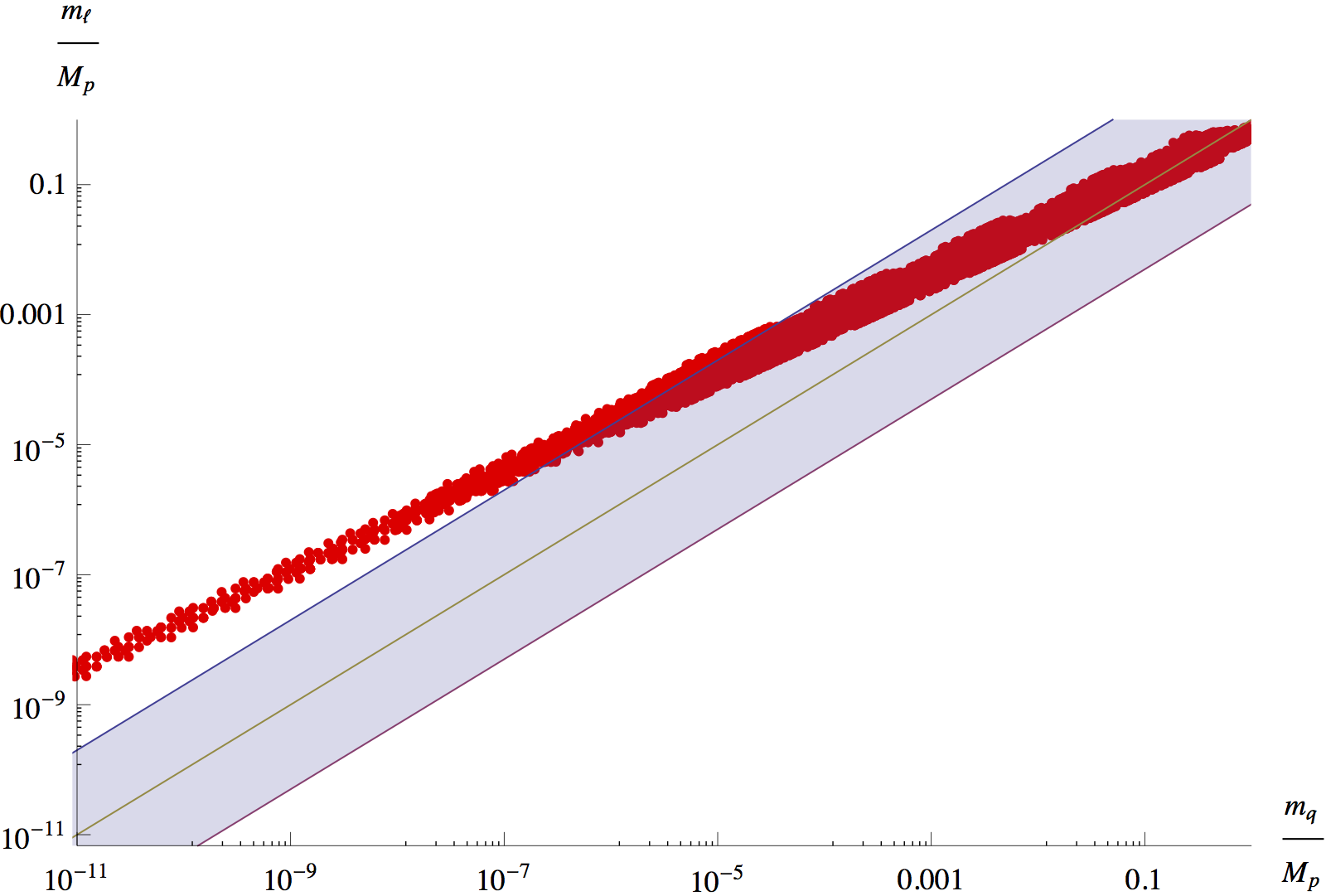}
\caption{In red, the masses of the fields $q$ and $\ell$ in the model of equations (\ref{Rmodel})--(\ref{SMcoupling}); the shaded area highlights the region where the two masses are comparable, that is, the same up to a factor of 20.}
\label{masses}
\end{center}
\end{figure}

To achieve gauge coupling unification, we require that the masses of $q$ and $\ell$ not be too different from each other.
We reintroduced all of the dimensionless couplings in equations (\ref{Rmodel}) and (\ref{SMcoupling}), varying them over a range of  values of order one.  We varied $\Lambda$ in coarse steps in a range from 
$10^{-4}M_p$ to $0.5 M_P$
and varied $N,\ F_1, \ F_2$ as well.  The results for the masses of $q$ and $\ell$ are shown in figure \ref{masses}. The masses of $q$ and $\ell$ are  approximately the same in a large region, in the range of $10^{13}-10^{18}$ GeV. \footnote{
In the model presented above, we almost always have $m_q<m_\ell$; this is because for most of our parameter space, $S_2>S_1$; then, the field coupling to $S_2$ will in general be heavier. One can easily build another model where the opposite behavior appears. If the couplings have the form
\beq
\frac{ S_1^2}M_p\bar q\bar q +\frac{S_2^3}{M_p^2}\bar\ell\ell
\eeq
we generically get $m_q > m_\ell$, while still being comparable on a range $10^{10}-10^{18}$ GeV.
}


\begin{figure}[tbhp]
\begin{center}
\includegraphics[width=7cm]{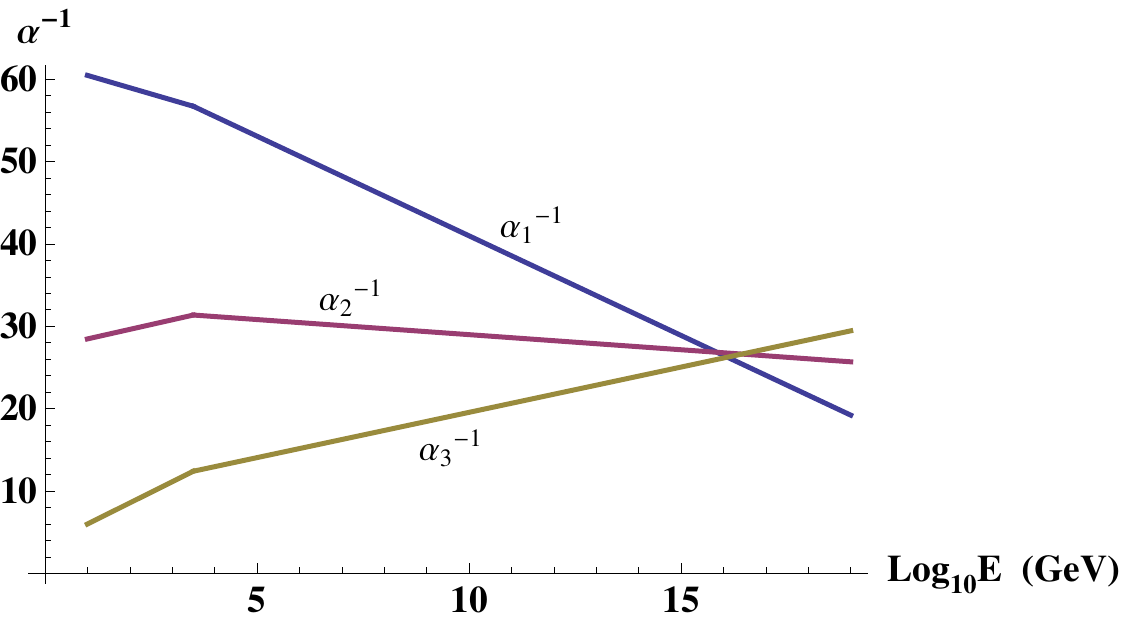}\ \ \
\includegraphics[width=7cm]{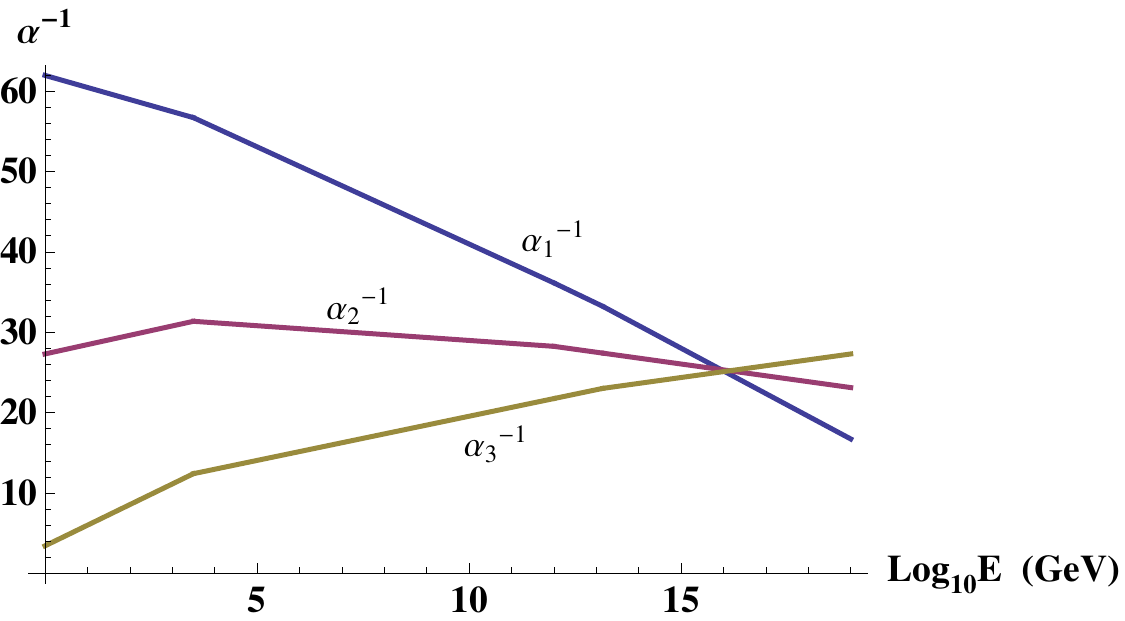}
\caption{Left: unification of the couplings in the simple MSSM, assuming the superpartners appear at the TeV scale. Right: unification of the couplings assuming a doublet appears at $10^{12}$ GeV and a triplet  at $10^{13}$ GeV.}
\label{gut}
\end{center}
\end{figure}

As an example of the effects on unification, we can take a point in figure \ref{masses} with $m_\ell=10^{-6}M_p=10^{12}$ GeV and $m_q=10^{13}$  GeV. For the MSSM, we will assume that the superpartners have a mass of 3 TeV. One sees from figure \ref{gut} that unification actually works better than in the MSSM: the $\mathcal O (2\%)$ mismatch is now at the permil level. Basically, this is because the $SU(2)$ coupling bends down earlier than the $SU(3)$ one.
These models should be embedded in more complete theories, e.g. as in the models of gauge and gravity mediation of \cite{bosedine}.  But it is clear that unification
of couplings is readily consistent with anomalous, approximate, discrete $R$ symmetries at low energies.

We conclude this section by discussing unification in more detail: it has been argued that non-universal axion couplings in the low-energy spoils gauge coupling unification, because the saxion expectation value sets the gauge coupling \cite{Chen:2012pi,Fallbacher:2011xg}. 
But already in typical string compactifications, there is no sense of unification in a semisimple group.  Instead, there are multiple moduli,
$S_i=s_i+ia_i$, with unification
arising (if at all) only because one modulus (call it $S_0$) with universal couplings to the various groups has an expectation value much larger 
 than the others (this is familiar in the heterotic string, and has been discussed more generally
in \cite{Ibanez:1991ai,Dienes:1996du} .  It is perfectly possible that  the axion-like fields $a_i$ cancel various non-universal anomalies while precision gauge unification is achieved with $1/g_i^2\sim \vev{s_0}$.

Even within more conventional grand unification, this possibility may arise;  it is not necessary that
the same linear combination of axions that cancels the high-energy anomaly couples universally to the Standard Model gauge groups.
As an existence proof, consider a field theory model in which unification is achieved in a product group: following the deconstructed models of \cite{wittendeconstruction}, we can consider a product group $SU(5)^4$. A field transforming in the bifundamental representation of the first two $SU(5)$'s can break $SU(5)^2$ to the diagonal subgroup $SU(3)_{diag}\times U(1)$, while another bifundamental in the last two $SU(5)$'s breaks them to $SU(2)_{diag}\times U(1)$. A linear combination of the two $U(1)$'s can be further broken by a bifundamental of the first and third $SU(5)$, leaving electromagnetic $U(1)_Y$ at low energy. If the various high-energy anomalies are cancelled by couplings of the form $f_iS_iW^2_{\alpha (i)}$, in the low energy the axions couple non-universally:
\beq
(f_1S_1+f_2S_2)W_{\alpha [SU(3)]}^2+(f_3S_3+f_4S_4)W_{\alpha [SU(2)]}^2+(f_1S_1+f_2S_2+f_3S_3+f_4S_4)W_{\alpha [U(1)]}^2
\eeq


 \section{Conclusions}
 \label{conclusions}
 
{\it If} supersymmetry has something to do with electroweak scale physics, discrete symmetries seem likely to play an important role.
Such symmetries, one expects, should be gauge symmetries, and should be free from anomalies.  Constraints from anomalies, then,
could provide interesting constraints on low energy model building.

This viewpoint, however, rests on strong assumptions about the underlying microscopic theory.  Arguably, some understanding of the microscopic theory is required
to determine the low energy constraints.  First, at best, without such knowledge, one can only impose constraints
involving anomalies connected to non-abelian gauge symmetries in the low energy theory.  The second complication
arises from the possibility
of Green-Schwarz cancellations.  We have seen in this paper that, already in simple Type II string compactifications,
there can be multiple scalar fields responsible for such cancellations, and as a result, {\it no constraints on the low
energy theory.}

One might counter that one should ignore the possibility of Green-Schwarz cancellations.  These require, after all, that one be in some
extreme limit of the moduli space.  Otherwise, the symmetry is badly broken at low energies and the question of anomalies
irrelevant.  On the other hand, the notion that one is in such a region is implicit in almost {\it all} discussions of string phenomenology,
where it is assumed that string couplings are weak, and (nearly as often) that compactification radii are large.
So it is difficult to put forth a doctrine, and it is interesting to explore a range of model building possibilities.  But one
needs to remember that, without knowledge of the microscopic theory or a model supported by experimental evidence,
that one can't put forward a reliably grounded set of rules.

In the case of discrete $R$ symmetries, there is a more immediate issue, in that any such symmetry is necessarily broken at a high energy scale.
While it is certainly possible that no fields with standard model quantum numbers gain mass
as a result of this breaking, we have seen that in rather simple models, this breaking can leave an {\it approximate} R symmetry at low energies, while at the same time
giving mass to combinations of fields which are chiral with respect to the symmetry.  The low energy theory is then anomalous.
So in the case of $R$ symmetries, it seems particularly hard to justify the imposition of anomaly constraints on the low energy theory.

More generally, the lessons of this paper apply to ``bottom up'', as opposed to ``top down'' model building.  It is certainly true that
many string constructions realize one or another set of possible anomaly constraints.  But we have seen that there are exceptions
to the various candidate anomaly constraints among well-studied string constructions, so from a purely low energy perspective,
none are compelling.


\begin{thebibliography}{12}

\bibitem{krausswilczek}
Lawrence~M. Krauss and Frank Wilczek.
\newblock {Discrete Gauge Symmetry in Continuum Theories}.
\newblock {\em Phys.Rev.Lett.}, 62:1221, 1989.

\bibitem{ibanezross}
Luis~E. Ibanez and Graham~G. Ross.
\newblock {Discrete gauge symmetry anomalies}.
\newblock {\em Phys.Lett.}, B260:291--295, 1991.

\bibitem{banksdinediscrete}
Tom Banks and Michael Dine.
\newblock {Note on discrete gauge anomalies}.
\newblock {\em Phys.Rev.}, D45:1424--1427, 1992.
\newblock Revised version.

\bibitem{yanagidaanomalies}
Kiichi Kurosawa, Nobuhito Maru, and T.~Yanagida.
\newblock Nonanomalous r symmetry in supersymmetric unified theories of quarks
  and leptons.
\newblock {\em Phys.Lett.}, B512:203--210, 2001.

\bibitem{ibanezuranga}
Luis~E. Ibanez, R.~Rabadan, and A.M. Uranga.
\newblock {Anomalous U(1)'s in type I and type IIB D = 4, N=1 string vacua}.
\newblock {\em Nucl.Phys.}, B542:112--138, 1999.

\bibitem{dinegraesser}
Michael Dine and Michael Graesser.
\newblock {CPT and other symmetries in string / M theory}.
\newblock {\em JHEP}, 0501:038, 2005.

\bibitem{gsw}
Michael~B. Green, J.H. Schwarz, and Edward Witten.
\newblock {Superstring Theory, Vol. 2: Loop Amplitudes, Anomalies and
  Phenomenology}.
\newblock 1987.

\bibitem{gepner}
Doron Gepner.
\newblock {Exactly Solvable String Compactifications on Manifolds of SU(N)
  Holonomy}.
\newblock {\em Phys.Lett.}, B199:380--388, 1987.

\bibitem{dinekehayias}
Michael Dine and John Kehayias.
\newblock {Discrete R Symmetries and Low Energy Supersymmetry}.
\newblock {\em Phys. Rev.}, D82:055014, 2010.

\bibitem{affleckdineseiberg} 
  I.~Affleck, M.~Dine and N.~Seiberg,
  Nucl.\ Phys.\ B {\bf 241}, 493 (1984).

\bibitem{wittendeconstruction} 
  E.~Witten,
  Deconstruction, G(2) holonomy, and doublet triplet splitting,
  hep-ph/0201018.


\bibitem{bosedine} 
  M.~Bose and M.~Dine,
  JHEP {\bf 1303}, 057 (2013)
  [arXiv:1209.2488 [hep-ph]].

\bibitem{Chen:2012pi}
Mu-Chun Chen, Maximilian Fallbacher, and Michael Ratz.
\newblock Supersymmetric unification and r symmetries.
\newblock {\em Mod.Phys.Lett.}, A27:1230044, 2012.

\bibitem{Fallbacher:2011xg} 
  M.~Fallbacher, M.~Ratz and P.~K.~S.~Vaudrevange,
  Phys.\ Lett.\ B {\bf 705}, 503 (2011)
  [arXiv:1109.4797 [hep-ph]].

\bibitem{Ibanez:1991ai} 
  L.~E.~Ibanez,
  In *Valencia 1991, Proceedings, Electroweak physics beyond the standard model* 351-372 and CERN Geneva - TH. 6342 (91/12,rec.Mar.92) 16 p
  [hep-th/9112050].

\bibitem{Dienes:1996du}
Keith~R. Dienes.
\newblock String theory and the path to unification: A review of recent
  developments.
\newblock {\em Phys.Rept.}, 287:447--525, 1997.

\end{thebibliography}

\end{document}